\def\spose#1{\hbox to 0pt{#1\hss}}
\def\lta{\mathrel{\spose{\lower 3pt\hbox{$\mathchar"218$}}
     \raise 2.0pt\hbox{$\mathchar"13C$}}}
\def\gta{\mathrel{\spose{\lower 3pt\hbox{$\mathchar"218$}}
     \raise 2.0pt\hbox{$\mathchar"13E$}}}

\documentclass[a4paper]{ISOStyle}
\usepackage{epsfig}

\begin{document}

\title{Far Infrared Emission from Elliptical Galaxies: NGC 4649, NGC 4472, and 
NGC 4636 }

\author{P. Temi\inst{1,2}, William G. Mathews\inst{3}, 
Fabrizio Brighenti\inst{3,4} \and J.D. Bregman\inst{1}} 
  
\institute{Astrophysics Branch, Nasa/Ames Research Center, MS 245-6, Moffett Field, CA 94035
  \and 
  SETI Institute, Mountain View, CA 94043 
 \and
University of California Observatories/Lick Observatory, Board of Studies
in Astronomy and Astrophysics, University of California, Santa Cruz, CA 94064
\and
Dipartimento di Astronomia, Universita' di Bologna, via Ranzani 1, Bologna 40127, Italy
}

\maketitle 

\begin{abstract}

\noindent
We present ISOPHOT P32 oversampled maps and P37/39 sparse maps,
of three bright elliptical galaxies in the Virgo Cluster. 
The maps reach the limiting
sensitivity of the ISOPHOT instrument at 60, 100, 170 and
200$\mu$m. Two elliptical galaxies show no emission at 
all far-IR ISOPHOT wavelengths at a level of few tens of mJy.  
The null detection provides a test of the evolution 
of dust in elliptical galaxies and its size distribution and 
composition.
\noindent
As previous studies have shown, in many elliptical galaxies both IRAS
and ISO have detected mid-IR excess 6-15 micron emission relative to
the stellar continuum indicating emission from circumstellar dust.
Under the assumption that these dusty outflows from evolving red giant
stars and planetary nebulae are continuously supplying dust to the
interstellar medium, we have computed the infrared luminosity at the
ISOPHOT bands appropriate for NGC4472. 
The null far-IR ISOPHOT observations exceed the far-IR flux 
expected from dust expelled from a normal old stellar population.

\end{abstract}

\section{INTRODUCTION}

In recent years the traditional view of elliptical galaxies as simple systems
of non interacting stars, devoid of interstellar matter, has radically changed.
Observations across the electromagnetic spectrum have demonstrated that the
interstellar medium (ISM) in elliptical galaxies contains substantial amounts
of cold gas and dust in addition to hot gas, the dominant component. Far-IR
emission detected by the IRAS satellite and improved optical imaging of elliptical
galaxies provide clear evidence for the presence of dust. However, estimates of
the total amount of dust, as well as its origin and spatial distribution, remain
uncertain and controversial.  Optical observations indicate dust masses that are
one order of magnitude  less than those inferred by IRAS.
We have recently embarked on a program to study the infrared emission from
early-type galaxies using the large database of observations  taken by the
ISO satellite. Our principal goals are to observe the  spatial location and
emission spectrum of dust throughout elliptical  galaxies, and with this
information, to determine the origin, evolution, and physical properties of
dust in massive elliptical galaxies. 

\section{OBSERVATIONS AND DATA REDUCTION}
\label{MSU_sec:tit}

\noindent
ISO provides a vast amount of data with
good spatial resolution in the mid and far-IR for a large number of
elliptical galaxies.
A good fraction of these  observations have been taken using 
ISOPHOT in several observing modes (Astronomical Observing Template), including 
the oversampled maps P32, the P37/38/39 sparse maps and P22 multi-filter 
photometry data. Here we present data from three very bright ellipticals 
located in Virgo Cluster: NGC4649, NGC4472, and NGC4636.

\noindent
NGC4636 has been observed in the P37/39 AOT mode at 60, 100, and 180 $\mu$m.
Observations were made in each filter with one on-source single point staring
mode and one off-source exposure. The background position was located $\sim10^\prime$ 
north-east  from the target, on blank sky position. Data reduction and calibration
were performed with the PIA 9.1 package (\cite{name:bib:gab97}). The reduction
included correction for the non linear response of the detectors, readout
deglitching, and linear fitting of the signal ramps. After resetting of all ramp
slopes and subtracting the dark current, the flux densities were extracted  using
the power calibration of the reference lamps. 
After sky subtraction the
on-source signal was corrected for the fraction of the point-spread
function included within the detector's field of view. A color correction
was applied assuming that the SED of the galaxy could be approximated by
a blackbody curve with temperature of 30 K. 
We tested the C100 detector response on a number of 
galaxies observed in the same observing mode as NGC4636. 
Integrated fluxes from 8 ellipticals were
compared to the IRAS values at  60 and 100 $\mu$m. Calibrated survey scans from the
IRAS satellite were extracted using the SCANPI software from IPAC. 
A very good linear correlation between the ISO and IRAS fluxes is seen at both wavelengths.

\begin{figure*}
  \begin{center}
    \epsfig{file=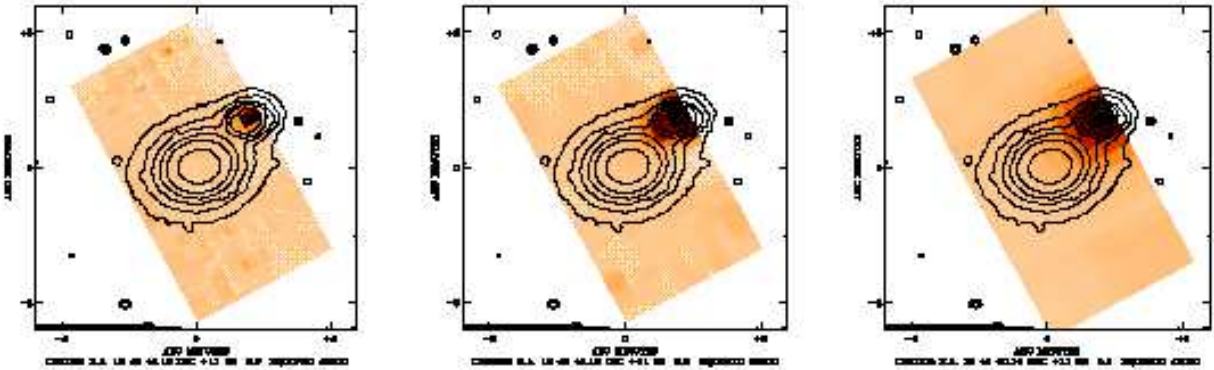, width=17cm}
  \end{center}
\caption{NGC4649 P32 maps at 60, 90, and 180 $\mu$m  respectively 
with optical contours superimposed. The galaxy is 
undetected at all wavelengths observed with ISOPHOT. The bright source
at the edge  of  map is its spiral companion NGC4647 and it is
prominent at all three wavelengths. The angular resolution of the
oversampled  map nicely resolves NGC4647 and shows that there is no
contribution from NGC4649 in this aperture centered on the
elliptical. 
At 180 $\mu$m the signal from the spiral galaxy extends almost to the
location of NGC4649 nucleus, but an 
analysis of the light profile shows that the
emission is clearly coming from the extended far-IR image of 
NGC4647.\label{name:fig:fig1}}
\end{figure*}

\begin{figure}[!hb]
  \begin{center}
    \epsfig{file=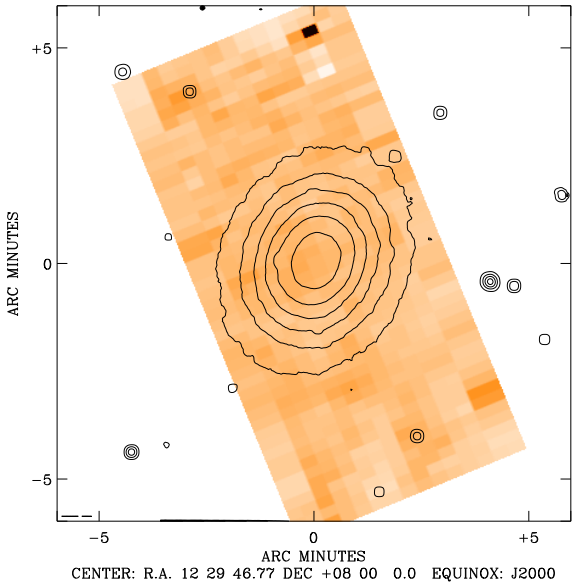, width=9.2cm}
  \end{center}
\caption{Optical  image (contours) superimposed to ISOPHOT 90$\mu$m map for
NGC4472 . Similar maps at 60 and 180$\mu$m show no detectable emission
from the bright elliptical.\label{name:fig:fig2}}
\end{figure}

\noindent
NGC4649 and NGC4472 have been observed in P32 mode in three broadband filters
at 60, 90 and 180 $\mu$m using the C100 and C200 detectors. The maps were obtained
by scanning the spacecraft Y and Z axes in a grid of 4$\times$4 (NGC4649) and
5$\times$4 (NGC4472) points. The focal plane
chopper was stepped at intervals of one-third of the detector pixel, at each
raster position, providing a sky sampling in the Y direction of 15" and 30"
for the C100 and C200 detectors. To reduce the P32 data we used a dedicated
software package developed at MPI Kernphysik in Heidelberg and supported by
the VILSPA ISO Data Center (\cite{name:bib:tuf03}). 
The new routines 
allow a proper correction for transients in PHT32 measurements. 
Both NGC4472 and
NGC4649 show no detectable emission at 60, 90, and 180 $\mu$m. The maps reach the
limiting sensitivity of the ISOPHOT instrument at a level of few tens of mJy
and the 3$\sigma$ upper limits are presented in Table 1. Figure 1 shows the 60, 90,
and 180$\mu$m maps for NGC4649. The bright spiral companion NGC4647, recorded 
at the edge of the maps, is well detected
at all three wavelengths and is responsible for the high flux detection, erroneously
attributed
to NGC4649, reported in literature from IRAS observations. In fact the angular 
resolution and beam size of the IRAS instruments were not able to disentangle 
the contribution of the two galaxies. Figure 2 show the 90 $\mu$m oversampled
map for NGC4472. 

\begin{table}[bht]
  \caption{ISOPHOT flux densities.}
  \label{MSU_tab:table}
  \begin{center}
    \leavevmode
    \footnotesize
    \begin{tabular}[h]{lccc}
      \hline \\[-5pt]
      \hline 
                   &             &  Flux (mJy)  &   \\[+5pt]
                   & NGC4649     & NGC4472      & NGC4636\\[+5pt]
      \hline \\[-5pt]
      $F_{60\mu m}$  &$ <$ 137$^{c}$ &$ <$ 147$^{c}$ & 187$\pm$ 57 \\
      $F_{90\mu m}$  &$ <$  85$^{c}$ &$ <$  99$^{c}$ & 491$\pm$ 64 \\
      $F_{180\mu m}$ &$ <$ 110$^{c}$ &$ <$  87$^{c}$ & 790$\pm$ 71 \\
      \hline \\
      \end{tabular}
  \end{center}
     $^ {c}$ {No detection, 3 $\sigma$ upper limit flux}
\end{table}

\section{A MODEL FOR NGC4472}
Mid-IR (6 - 15$\mu$m) ISO observations of NGC4472 detect strong 
dust emission that can be understood as emission 
from circumstellar dust from mass-losing AGB stars 
at or near the 
globally expected rate for this galaxy (\cite{name:bib:ath02}). 
Since this dust should emit in the far-IR when it 
becomes interstellar, it may be significant that NGC4472 is not 
detected at $\lambda \geq 25\mu$m 
either by ISO or IRAS (\cite{name:bib:kna89}).
We now briefly describe an estimate of 
the far infrared emission from NGC4472, assuming that the 
grains are dispersed in the hot interstellar gas before sputtering 
commences.
First we calculate 
the temperature $T_d(r,a)$ of dust grains of radius
a (in $\mu$m) at galactic radius $r$ and
then determine the grain emissivity and luminosity
for a distribution of initial grain sizes.
Since the 9.7$\mu$m silicate feature 
is seen in emission in NGC4472, we 
assume that all grains have properties similar to astronomical 
silicates (\cite{name:bib:lao93}).\\
The grain temperature 
is determined by both absorption of starlight 
and by electron-grain collisions:
\begin{eqnarray}
\lefteqn
{\int_0^{\infty} 4 \pi J_*(r,\lambda) Q_{abs}(a,\lambda)
\pi a^2 d \lambda
+ 4 \pi a^2 (1/4) n_e \langle v_e E_e \rangle  
\tau(a)}\nonumber\\
& &= 4 \pi a^2 \sigma_{SB} T_d(r,a)^4  
\langle Q_{abs} \rangle (T_d, a)
\end{eqnarray}
although we consider each type of heating separately.
The mean intensity of starlight  at wavelength
$\lambda$ (in $\mu$m) is related to the B-band intensity 
$J_*(r,\lambda) = J_{\lambda B}(r) \phi(\lambda)$ 
and the SED $\phi(\lambda)$ taken from \cite*{name:bib:tsa95}. 
$J_{\lambda B}(r)$ is found by integrating over a de Vaucouleurs 
stellar profile with a ``nuker'' core of slope $\rho_* \propto
r^{-0.95}$ within $0.031 R_e$ (\cite{name:bib:fab97}). 
At a distance $d = 17$ Mpc NGC4472 has an effective radius 
$R_e = 8.57$ kpc and total stellar mass $M_{*,t} = 7.26 \times
10^{11}$ $M_{\odot}$ assuming 
$M_{*,t}/L_B = 9.2$ (\cite{name:bib:van91}).
The second term on the left represents heating 
by e$^-$-grain collisions including a correction 
$\tau(a)$ for small grains which do not completely stop the 
electrons (\cite{name:bib:dwe86}).
For astronomical silicate grains of radius less than $\sim 1\mu$m 
we can use the approximations 
$Q_{abs} \approx a \psi(\lambda)$ and 
$\langle Q_{abs} \rangle \approx 1.35 \times 10^{-5} T_d^2 a$. 
For starlight heating the grain temperature 
$T_d(r) = 44 J_{\lambda B}^{1/6}$ K 
is independent of $a$ and varies from 
26 K at 1 kpc to 16 K at 10 kpc. 
For e$^-$-grain heating the grain temperature 
$T_d = 0.662 (n_e T^{3/2} \tau / a)^{1/6}$ K 
varies as $a^{-1/6}$ for $a > 0.04\mu$m where 
$\tau \approx 1$ but is independent 
of grain radius for $a < 0.04\mu$m where $\tau \approx 33 a$.
Figure 3 shows the calculated dust temperature profile for the
starlight heating and the e$^-$-grain heating. For NGC 4472 the
collisional heating dominates over the radiative heating for
grain sizes $a \lta 0.3$ $\mu$m.
  
\begin{figure}[!h]
  \begin{center}
    \epsfig{file=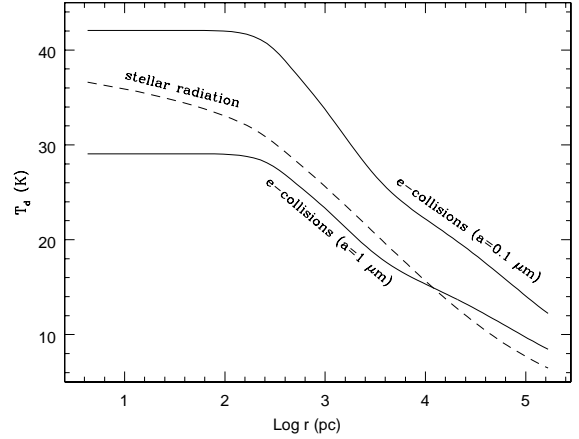, width=9.2cm}
  \end{center}
\caption{Temperature of dust grains due to starlight heating only
(dashed lines), and due to electron-grain collisions only (heavy 
solid line for grain size $a=1$ $\mu$m; thin solid line for grain size
$a=0.1$ $\mu$m). For grains with size $a\lta 0.3$ $\mu$m the collisional
heating is larger than the radiative one. 
\label{name:fig:fig3}}
\end{figure}

\noindent 
If grains are expelled from the stars with
an initial MRN size distribution
$$S(a,r)  =  S_o(r) a^{-s}$$
where $s = 3.5$ and $a$ is in $\mu$m
(\cite{name:bib:mat77}), 
then the steady state grain size distribution
in the hot gas is 
\begin{equation}
N(r,a) = \left| {da \over dt} \right|^{-1}
{S_o(r) \over (s - 1) } a^{1-s}~~~a \leq a_{max}.
\end{equation}
The coefficient 
\begin{equation}
S_o(r) = {3 \delta \alpha_* \rho_* \over 
4 \pi \rho_g 10^{-12}} { (4 - s) \over a_{max}^{4-s}}
\end{equation}
depends on the rate that the old stellar 
population ejects mass,
$\alpha_* = 4.7 \times 10^{-20}$ s$^{-1}$,
the initial dust to gas mass ratio scaled
to the stellar metal abundance in NGC4472,
$\delta = (1/150)z_m(r)$ 
and the density of silicate grains,
$\rho_g = 3.3$ gm cm$^{-3}$. 
The grain sputtering rate 
\begin{equation}
{d a \over dt} = - fn_p 3.2 \times 10^{-18}
\left[ \left( {2 \times 10^6~{\rm K} \over T} \right)^{2.5} + 1
\right]^{-1}~~~\mu{\rm m}~{\rm s}^{-1}
\end{equation}
is a fit to the rate of \cite*{name:bib:dra79} 
with an additional correction 
$f = f(a,T) \geq 1$ to account 
for the enhanced (isotropic) sputtering from small grains 
(\cite{name:bib:jur98}).
The hot gas temperature $T$ and proton density $n_p$ 
for NGC4472 are 
taken from \cite*{name:bib:bri97}. 

\noindent
The infrared emissivity from grains at galactic radius $r$ is
\begin{eqnarray}
j(r,\lambda) = {1 \over 4 \pi} \int_{0}^{a_{max}}
da N(r,a) 4 \pi a^2 10^{-8} Q_{abs}(a,\lambda) \nonumber \\
\times \pi B(r,a,\lambda)
\end{eqnarray}
where $B(T_d(r),a,\lambda)$ is the Planck function.

\noindent
The total galactic luminosity at wavelength $\lambda$ is
\begin{equation}
L_{\lambda} = \int 4 \pi j(r,\lambda) (4/3) \pi d(r^3)
\end{equation}
and the observed flux at $\lambda_{\mu}$ is 
\begin{equation}
F_{\nu} = 10^{22} {L_{\lambda} \over 4 \pi d^2}
{\lambda_{\mu{\rm m}}^2 \over c}~~~{\rm mJy}.
\end{equation}

Figure 4 compares the computed far-IR spectra for the cases
of stellar radiation heating and collisional heating with the 
three ISO flux upper limits.
The total IR emission (not shown in the figure) is the 
sum of the contribution of the two heating sources
for each value of $a_{max}$.
Grains with an initial 
MRN distribution having $a_{max} < 1\mu$m 
are consistent with the null observations 
of NGC4472 in the far-IR.
Grains with radii less than about $0.003\mu$m experience 
stochastic temperature excursions, but this complication 
should not greatly change our estimated flux in 
Figure 4.

\begin{figure}[!h]
  \begin{center}
    \epsfig{file=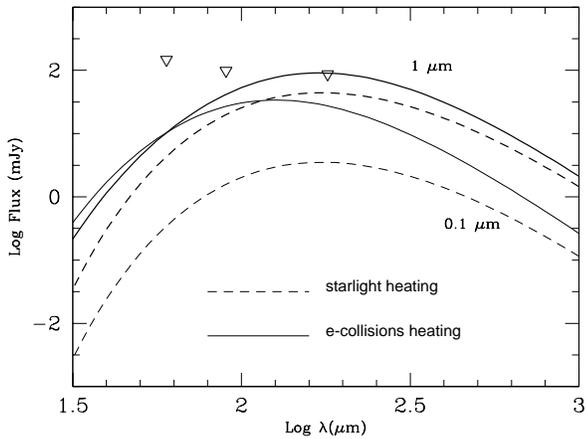, width=9.2cm}
  \end{center}
\caption{ Computed Far-IR flux in mJy from NGC4472 
calculated 
for $a_{max} = 1\mu$m (heavy lines) and 
$a_{max} = 0.1\mu$m (light lines).
Solid lines show the grain emission based only 
on electron-grain heating and dashed lines show 
the emission from grains that are heated only by 
stellar radiation. 
The three ISO upper limits are shown with open 
triangles. \label{name:fig:fig4}}
\end{figure}

\section{RESULTS}

\noindent
We have assumed that the circumstellar dust observed
in the mid-IR from NGC4472 (\cite{name:bib:ath02})
ultimately becomes interstellar and is
exposed to the mean galactic starlight and 
to bombardment by thermal electrons in the hot
$\sim 1$ keV gas.
Grains receive somewhat more 
energy from electrons than from stellar photons.
The calculated far-IR spectrum peaks at
$\lambda \sim 180$ $\mu$m, a spectral region not accessible to IRAS,
but well covered by ISO.
However, the computed fluxes are below the observed upper limits
if $a \lta 0.5$ $\mu$m. Thus
the ISO upper limits do not strongly constrain the 
properties of dust in NGC4472. We are currently developing a model
for the far-IR emission for NGC4636 (\cite{name:bib:tem02}).
This galaxy shows higher far-IR fluxes (Table 1) and should allow
us to put strong constraints on the origin and the evolution
of the intestellar dust.

We thank NASA and NSF for grants that funded this research.

\end{document}